\documentclass[letterpaper,twocolumn,10pt]{article}
\usepackage{usenix}
\usepackage{amsthm}

\usepackage{tikz}
\usepackage{amsmath}

\usepackage{tikz}
\usepackage{amsmath}

\usepackage{graphicx, multirow}
\usepackage[T1]{fontenc}
\usepackage{semantic}

\usepackage{tabularx, multirow, rotating} 
\usepackage{subfig} 
\usepackage{algorithm} 
\usepackage{algpseudocode} 
\usepackage{ulem} 
\usepackage{color} 

\usepackage{marvosym}

\usepackage[most]{tcolorbox}
\usepackage{listings} 
\lstdefinestyle{Oracle}{basicstyle=\ttfamily,
                        keywordstyle=\lstuppercase,
                        emphstyle=\itshape,
                        showstringspaces=true,
                        }
\makeatletter
\newcommand{\lstuppercase}{\uppercase\expandafter{\expandafter\lst@token
                           \expandafter{\the\lst@token}}}
\newcommand{\lstlowercase}{\lowercase\expandafter{\expandafter\lst@token
                           \expandafter{\the\lst@token}}}
\makeatother

\usepackage{tikz}

\usepackage{tikz}
\newcommand{\ballnumber}[1]{\tikz[baseline=(myanchor.base)] \node[circle,fill=.,inner sep=1pt] (myanchor) {\color{-.}\bfseries\footnotesize #1};}

\newif\ifboldnumber

\algrenewcommand\alglinenumber[1]{%
  \footnotesize\ifboldnumber\bfseries\fi\global\boldnumberfalse#1:}

\usepackage{color, colortbl}
\definecolor{Gray}{gray}{0.9}
\definecolor{LightCyan}{rgb}{0.88,1,1}
\definecolor{Azul}{rgb}{0.16, 0.32, 0.75}
\usepackage[first=0,last=9]{lcg}

\usepackage{colortbl}
\usetikzlibrary{calc}
\usepackage{zref-savepos}

\newcounter{NoTableEntry}
\renewcommand*{\theNoTableEntry}{NTE-\the\value{NoTableEntry}}

\usepackage{enumitem}
\usepackage{amsfonts}

\usepackage{tablefootnote}
\usepackage{wrapfig}
\usepackage{float}
\usepackage{multirow}
\usepackage{graphicx}
\usepackage{algorithm}
\usepackage{algpseudocode} 
\usepackage{amsmath}
\usepackage{pgf-pie}
\usepackage{alphalph}
\usepackage{xspace}

\usepackage{url}

\usepackage{booktabs}
\usepackage{stackrel}

\usepackage{tikz}

\usepackage{tikz}
\newcommand{\ignore}[1]{}
\usepackage{bm}
\usepackage{pifont}
\newcommand{\cmark}{\ding{51}}%
\newcommand{\xmark}{\ding{55}}%

\usepackage{todonotes}

\usepackage{graphicx, multirow}

\usepackage{tabularx, multirow, rotating} 
\usepackage{subfig} 
\usepackage{algorithm} 
\usepackage{algpseudocode} 
\usepackage{ulem} 
\usepackage{color} 

\usepackage{listings} 
\lstdefinestyle{Oracle}{basicstyle=\ttfamily,
                        keywordstyle=\lstuppercase,
                        emphstyle=\itshape,
                        showstringspaces=true,
                        }

\usepackage{tikz}

\usepackage{tikz}

\newif\ifboldnumber
\algrenewcommand\alglinenumber[1]{%
  \footnotesize\ifboldnumber\bfseries\fi\global\boldnumberfalse#1:}

\usepackage{color, colortbl}
\definecolor{Gray}{gray}{0.9}
\definecolor{LightCyan}{rgb}{0.88,1,1}
\usepackage[first=0,last=9]{lcg}

\usepackage{colortbl}
\usetikzlibrary{calc}
\usepackage{zref-savepos}

\renewcommand*{\theNoTableEntry}{NTE-\the\value{NoTableEntry}}

\begin{document}

\date{}


\title{More to Extract: Discovering MEV by Token Contract Analysis}


\author{
{\rm Jiaqi Chen}\\
Syracuse University\\
\texttt{jchen217@syr.edu}
\and
{\rm Yuzhe Tang \textsuperscript{\Letter}}\\
Syracuse University\\
\texttt{ytang100@syr.edu}
\and
{\rm Yue Duan}\\
Singapore Management University\\
\texttt{yueduan@smu.edu.sg}
}

\maketitle
\begingroup
\renewcommand\thefootnote{\Letter}
\footnotetext{Yuzhe Tang is the corresponding author.}
\endgroup

\providecommand{\cMEV}{{\sc tMEV}\xspace}
\providecommand{\staticname}{{\sc tSCAN}\xspace}
\providecommand{\dynamicname}{{\sc tSEARCH}\xspace}
\providecommand{\csearchr}{{\sc tSEARCH}\xspace}
\providecommand{\qone}{{\sc lend}\xspace}
\providecommand{\qtwo}{{\sc swap}\xspace}
\providecommand{\qthree}{{\sc addL}\xspace}

\begin{abstract}
This paper tackles the discovery of tMEV, that is, the \underline{M}aximal \underline{E}xtractable \underline{V}alue on blockchains that arises from \underline{T}oken smart contracts.
This scope differs from the existing MEV-discovery research, which analyzes application-layer contracts or attacker contracts, but ignores the wide and diverse range of token contracts.

This paper presents a pipeline of techniques for tMEV discovery, including tSCAN, a static analysis tool for identifying non-standard supply-control functions in token contracts, and tSEARCH, a searcher that uncovers profitable tMEV opportunities by generating, refining, and solving token-specific constraints.

By replaying real-world transactions, this paper demonstrates both the profitability of tMEV strategies and existing searchers' unawareness of them: the proposed tSEARCH extracts $10\times$ more profit than observed MEV activity on Ethereum.
The practicality of tMEV searching is demonstrated through a prototype built on Slither, showing high effectiveness with low performance overhead.

\ignore{
This paper explores a previously unexamined source of blockchain-extractable value (MEV), termed \cMEV, which arises from the composition of token-supply-control (TSC) mechanisms and price-insensitive exchanges such as lending services and Uniswap V3/V4.

To enable efficient \cMEV discovery, we develop tSCAN, a static analysis tool for identifying non-standard TSC functions in token contracts, and tSEARCH, a stateful MEV searcher that uncovers multi-step profitable opportunities by incrementally generating and solving execution constraints.

By replaying real-world transactions, this paper demonstrates both the profitability of \cMEV strategies and existing searchers' unawareness to them: the proposed \dynamicname{} extracts $10\times$ more profit than observed MEV activity on Ethereum. 
The practicality of \cMEV searching is demonstrated through a prototype built on Slither, showing high effectiveness with low performance overhead.
}

\ignore{
This paper tackles the discovery of token-driven maximal extractable value (tMEV) on Ethereum -- a largely overlooked area in the existing MEV research, which has focused on analyzing application-layer smart contracts instead of tokens.

We present tSCAN, a static smart-contract analysis tool for token Supply Control ANalysis, and tSCAN-s, a runtime system for dynamic tMEV searching in real time.
}

\ignore{
In Ethereum, maximal extractable value (MEV) impacts both the financial risk of crypto-assets in DeFi contracts and the security of blockchain consensus.  
Discovering new MEV patterns, beyond well-known and thus highly competitive ones, is crucial, as they offer greater profit potential.  
While existing research primarily analyzes application-layer contracts, it largely overlooks token contracts. This paper fills that gap by statically analyzing token smart contracts to uncover new profitable MEV opportunities.
}
\end{abstract}


\definecolor{mygreen}{rgb}{0,0.6,0}
\lstset{ %
  backgroundcolor=\color{white},   
  basicstyle=\scriptsize\ttfamily,        
  breakatwhitespace=false,         
  breaklines=true,                 
  captionpos=b,                    
  commentstyle=\color{mygreen},    
  deletekeywords={...},            
  escapeinside={\%*}{*)},          
  extendedchars=true,              
  keepspaces=true,                 
  keywordstyle=\color{blue},       
  language=Java,                 
  morekeywords={*,...},            
  numbers=left,                    
  numbersep=5pt,                   
  numberstyle=\scriptsize\color{black}, 
  rulecolor=\color{black},         
  showspaces=false,                
  showstringspaces=false,          
  showtabs=false,                  
  stepnumber=1,                    
  stringstyle=\color{mymauve},     
  tabsize=2,                       
  title=\lstname,                  
  moredelim=[is][\bf]{*}{*},
}

\colorlet{blue}{black} \colorlet{violet}{black} \colorlet{green}{black}  
\colorlet{teal}{black}

\lstdefinestyle{mystyle}{
    backgroundcolor=\color{backcolour},   
    commentstyle=\color{codegreen},
    keywordstyle=\color{codepurple},
    numberstyle=\numberstyle,
    stringstyle=\color{blue},
    basicstyle=\footnotesize\ttfamily,
    breakatwhitespace=false,
    breaklines=true,
    captionpos=b,
    keepspaces=true,
    numbers=left,
    numbersep=10pt,
    showspaces=false,
    showstringspaces=false,
    showtabs=false,
}
\lstset{style=mystyle}

\newcommand\numberstyle[1]{%
    \footnotesize
    \color{codegray}%
    \ttfamily
    \ifnum#1<10 0\fi#1 |%
}

\definecolor{mygreen}{rgb}{0,0.6,0}
\lstset{ %
  backgroundcolor=\color{white},   
  basicstyle=\scriptsize\ttfamily,        
  breakatwhitespace=false,         
  breaklines=true,                 
  captionpos=b,                    
  commentstyle=\color{mygreen},    
  deletekeywords={...},            
  escapeinside={(*}{*)},          
  extendedchars=true,              
  keepspaces=false,                 
  keywordstyle=\color{blue},       
  language=Java,                 
  morekeywords={*,...},            
  numbers=left,                    
  numbersep=5pt,                   
  numberstyle=\scriptsize\color{black}, 
  rulecolor=\color{black},         
  showspaces=false,                
  showstringspaces=false,          
  showtabs=false,                  
  stepnumber=1,                    
  stringstyle=\color{purple},     
  tabsize=2,                       
  title=\lstname,                  
  moredelim=[is][\bf]{*}{*},
}

\definecolor{codegreen}{rgb}{0,0.6,0}
\definecolor{codegray}{rgb}{0.5,0.5,0.5}
\definecolor{codepurple}{HTML}{C42043}
\definecolor{backcolour}{HTML}{F2F2F2}
%
%
%

\section{Introduction}

In public blockchains, notably Ethereum, the order in which transactions are included in a newly produced block can differ from the order in which they are submitted or propagated. This enables a searcher account, upon observing unconfirmed victim transactions, to submit their own transactions to manipulate the block-inclusion order and extract value from the victim. This class of financial attacks, known as maximal extractable value (MEV), increases the risk to crypto-assets such as Ether and tokens managed by decentralized finance (DeFi) smart contracts. MEV also plays a critical role in the incentive compatibility and consensus security of the blockchain itself, as it directly affects validator revenue.

\noindent{\bf Related works \& problem}:
In the literature, passive measurement studies~\cite{DBLP:conf/sp/QinZG22,DBLP:conf/sp/DaianGKLZBBJ20,DBLP:conf/uss/McLaughlinKV23,DBLP:conf/ccs/LiLHLWNYCC23,DBLP:conf/uss/TorresCS21,DBLP:conf/www/WangCWZDW22,DBLP:conf/uss/0002LM0023,DBLP:conf/uss/QinCZLSG23} have been conducted to rediscover MEV activities which have already occurred on chain.

In contrast, {\it proactive MEV discovery} aims to identify new MEV patterns before they appear on-chain, offering a first-mover advantage in profitability. Along this line, works such as DefiPoser~\cite{DBLP:conf/sp/ZhouQCLG21} and Lanturn~\cite{DBLP:conf/ccs/BabelJ0KKJ23} find function arguments that maximize profit for a given MEV transaction sequence using protocol-level models rather than concrete implementations. For example, different DEX implementations, whether Uniswap V2, V3, or others, are abstracted into the same set of constraints derived from a reference DEX model. Such techniques overlook MEV opportunities that arise from the concrete smart contracts implementing these DeFi protocols.


This work aims at the proactive discovery of {\it concrete MEV} opportunities. A straw-man approach is to analyze smart contracts across different layers of a target protocol, such as the application and token layers.
Such compositional analysis, however, greatly expands the search space and is computationally expensive. As a result, existing compositional analysis techniques are applicable to either  protocol analysis~\cite{DBLP:conf/sp/BabelDKJ23} or concrete smart-contract analysis; the latter entails efficiency through abstractions, leading to underapproximation and missing certain classes of MEV opportunities. For example, Foray~\cite{DBLP:conf/ccs/WenLSC0024} and similar works~\cite{zhang2025followingdevilsfootprintrealtime} construct token flow graphs by modeling DeFi operations from standard token-function calls; MEV opportunities are identified as cycles on the graph. Foray could miss MEV that arises from non-cycle structures or non-standard token functions. Nyx~\cite{DBLP:conf/sp/ZhangZSLWLZC24} discovers MEV by statically analyzing smart contracts and taint tracking data dependencies, such as price variables. However, this approach misses MEV derived from dependencies that only manifest at runtime. See \S~\ref{sec:relatedwork:1} for detailed analysis and evaluation.

A motivating example concretizing the above observation arises from AMPL~\cite{me:ampl:token}, a widely used token that adjusts token supply and updates {\it all} accounts' balances via its \texttt{rebase} function.
Consider a transaction sequence:
$tx_0$: $dY1=[a, q]$.\texttt{swapXY$(dX)$},
$tx_1$: $[b, Y]$.\texttt{rebase$()$},
$tx_2$: $[a, q]$.\texttt{swapYX$(dY2)$}.
Here, searcher $a$ swaps her initial Token $X$ into Token $Y$ with Pool contract $q$ (in $tx_0$), receives additional $Y$ via Token contract $Y$'s \texttt{rebase} event (in $tx_1$), and then swaps $Y$ back to $X$ (in $tx_2$) at the {\it original, unchanged price}. Thus, the additional amount of Token $Y$ (i.e., $dY2-dY1$) creates a positive profit to searcher $a$, as confirmed in the detailed profitability analysis in \S~\ref{sec:moex:cmev} (see Example D1+).

Existing MEV-discovery tools either analyze application-layer smart contracts (e.g., DEX pools as analyzed by Nyx~\cite{DBLP:conf/sp/ZhangZSLWLZC24}, Foray~\cite{DBLP:conf/ccs/WenLSC0024}, and DeFiTainter~\cite{DBLP:conf/issta/KongCWJZ23}) or attacker contracts (e.g., SmartCAT~\cite{DBLP:conf/uss/ZhangHHMW25}), they do not capture the MEV in the above example that arises from non-standard token implementations, such as in \texttt{rebase} function. Empirically, we validate Nyx's incapability of detecting our MEV (\S~\ref{sec:relatedwork:1}).

\begin{table}[!htbp]
\centering
\caption{Comparing the scopes of different smart-contract analysis for profit discovery. Token/Pool/Attacker refers to the smart contracts being analyzed.}
\label{tab:scope}
\begin{tabularx}{0.375\textwidth}{X|ccc}
Name & Token & Pool & Attacker \\
\hline
\hline
Nyx~\cite{DBLP:conf/sp/ZhangZSLWLZC24}, 
 & 
\multirow{3}{*}{\xmark} & 
\multirow{3}{*}{\cmark} & 
\multirow{3}{*}{\xmark} 
\\
Foray~\cite{DBLP:conf/ccs/WenLSC0024},
&&& \\
DeFiTainter~\cite{DBLP:conf/issta/KongCWJZ23}
&&& \\ 
\hline
SmartCAT~\cite{DBLP:conf/uss/ZhangHHMW25}
 & \xmark & \xmark & \cmark \\
\hline
This work 
 & \cmark & \xmark & \xmark \\
\end{tabularx}
\end{table}

\noindent{\bf Approach}:
This work bridges the gap in the existing literature and discover the MEV opportunities by analyzing token smart contracts, as shown in Table~\ref{tab:scope}.
Our MEV-discovery pipeline covers both static analysis and dynamic search.
First, we detect token-supply-control (TSC) functions in token smart contracts, defined by generalizing token \texttt{rebase}, TSC are functions that adjust balances across multiple token holders.
We propose static token-contract analysis, dubbed \staticname{}, that extracts data- and control-dependencies, and searches for reachable execution paths that update balances across multiple accounts, as in TSC functions.

Second, given the discovered TSC functions, we plug them into constraints derived from the observed \cMEV patterns, namely the composition of TSC and price-insensitive pools as aforementioned.
The solvability of these constraints confirms the profitability of the corresponding \cMEV patterns.
In this work, we manually instantiate price-insensitive pools from representative DeFi applications, including most lending protocols (e.g., Aave, which typically uses fixed prices) and popular AMM protocols such as Uniswap V3 and V4 (while other AMMs use sensitive prices).

We turn this design into a functional transaction-processing pipeline integrated with Ethereum mempools and MEV searching.
In the offline phase, the pipeline runs the \staticname{} token analysis and produces ``static'' constraints with runtime arguments left symbolic.
In the online phase, the pipeline runs a customized MEV searcher, dubbed \dynamicname{}, which monitors submitted unconfirmed transactions, uses them to concretize the static constraints with runtime information, and solves them into transaction sequences for \cMEV extraction.

\vspace{0.1in}
\noindent{\bf
Evaluation}: 
We implemented \staticname{} on top of Slither~\cite{me:slither} and developed \dynamicname{} in a Python-based runtime system that interacts with Ethereum via RPC. Using transactions from Ethereum mainnet, we evaluate the profitability of \cMEV, characterize real-world TSC tokens, and assess the practicality of our MEV search tool.

First, our measurement shows that real-world searchers have low awareness of the full spectrum of \cMEV strategies, resulting in profit-suboptimal activity on Ethereum. A wide range of highly profitable strategies, including sandwich-based \cMEV, those instantiated by non-standard price-insensitive pools, and \cMEV locking and unlocking, are entirely absent from observed transactions.
A what-if analysis further shows that our \dynamicname{} can effectively explore optimal \cMEV strategies, extracting $\$2.28*10^6$ from $7,030$ deployed pools, achieving $10\times$ higher profit than current mainnet activity on the same pools.

Second, we identify $5,434$ TSC tokens from $22,279$ collected token contracts. These tokens exhibit a wide variety of mechanisms, including explicit triggers via one or multiple function calls, implicit triggers through the change of blockchain state, and updates to the balances of one or multiple accounts.

Third, \staticname{} achieves a low false-positive rate ($1/53 = 1.8\%$) and a low false-negative rate ($1/247 = 0.40\%$) on our manually labeled ground-truth dataset.

\noindent{\bf Contributions} of this paper are as follows.

\vspace{2pt}\noindent$\bullet$\textit{
New problem}: 
This paper tackles the open research problem of discovering MEV strategies arising from the previously unstudied composition of {\bf non-standard tokens} and {\bf price-insensitive exchanges}, termed \cMEV. In contrast, prior MEV research has focused on price-sensitive exchanges and standard token functions.

\vspace{2pt}\noindent$\bullet$\textit{
New approach}:
This paper presents \staticname, a static analysis tool for detecting non-standard token-supply-control (TSC) functions. It also introduces \dynamicname, an efficient \cMEV searcher that tracks multiple victim transactions. 
Together, the searcher system is designed to discover \cMEV arising from the composition of detected TSC functions and price-insensitive exchanges, including lending services and lesser-known cases like token swaps and inactive liquidity additions/removals in Uniswap V3/V4.

\vspace{2pt}\noindent$\bullet$\textit{
Real-world results}:
By replaying real-world transactions, this paper demonstrates both the profitability of \cMEV strategies and existing searchers' unawareness of them: the proposed \dynamicname{} extracts $10\times$ more profit than observed MEV activity on Ethereum. 
The practicality of \cMEV searching is demonstrated through a prototype built on Slither, showing high effectiveness with low performance overhead.

\section{Background}
\label{sec:motivating:1:examples}

\noindent{\bf 
AMM pool}:
An Automated Market Maker (AMM) pool supports token swaps: A trader can call a pool's function $\texttt{swapXY}(X^a)$ to transfer $X^a$ tokens of $T_X$ to the pool while receiving $Y^a$ tokens of $T_Y$ from the pool. The pool enforces a certain invariant to calculate the exchange price (i.e., $\frac{Y^a}{X^a}$). One common invariant is the constant product of pool token reserves. Specifically, given that the pool initially owns $X^p_0$ tokens of $T_X$ and $Y^p_0$ tokens of $T_Y$, a constant-product AMM pool or so-called CPMM, enforces the following invariant to decide the token price: 

$$ 
(X^p_0+X^a)(Y^p_0-Y^a) = X^p_0\cdot{}Y^p_0
$$

\noindent{\bf 
Token supply control (TSC)}:
In a token smart contract, circulating token supply, or supply for short, refers to the number of tokens issued and being circulated. 
Typically, the token supply is created when the token smart contract is deployed on-chain (i.e., initialized in the token constructor). After that, the token supply may be posthumously adjusted, for instance, in order to control the token's price. We refer to the posthumous token supply adjustment as supply control. A token transfer across different accounts does not change the token supply. 

Unlike token transfers, whose interfaces are standardized (e.g., functions \texttt{transfer} and \texttt{transferFrom} in ERC20), supply-control functions are not standardized.
For instance, \texttt{mint} and \texttt{burn} are not part of standard ERC-20 or ERC-777 interfaces. However, they are commonly implemented in popular token libraries and protocols such as OpenZeppelin~\cite{me:erc20:api} and Uniswap LP tokens~\cite{me:uniswapLP}, typically using custom interfaces. These functions often differ in accessibility. Notably, \texttt{mint} is typically defined as an internal method and invoked through contract-specific public functions that enforce access control conditions.

\begin{table}[!htbp] 
\caption{Notations (\# means number).}
\label{tab:notations}
\centering
\small
\begin{tabularx}{0.48\textwidth}{l X l X}
 & Meaning &  & Meaning \\ 
\midrule
$p$ & price-sensitive pool & $q$ & price-insensitive pool \\
$st_i$ & blkchain state before $tx_i$ & $A^Y$ & all accounts holding $T_Y$
\\
$Y^a_i$ & \multicolumn{3}{l}{\# tokens $T_Y$ owned by account $a$ in state $st_i$}
\\
\end{tabularx}
\end{table}

Table~\ref{tab:notations} summarizes the notions used in this paper. 
We use a number $i$ to identify a transaction; note that the index may or may not refer to the position of the transaction in the globally ordered transaction history (when the transaction is confirmed). Given a transaction $tx_i$, let $st_i$ denote its pre-state, that is, the set of blockchain-native variables (e.g., \texttt{block.number}) and smart-contract storage variables right before $tx_i$ is executed. 
Given account $a$ and blockchain state $st_i$, we use $Y^a_i$ to denote the number of tokens held by account $a$ in token contract $T_Y$ as of state $st_i$. 
$p$ denotes an AMM pool, and $q$ denotes an AMM pool whose price is insensitive to the pool's token balance.

\section{Threat Model}
\label{sec:threatmodel}

In our threat model, there exist two tokens $T_X$ and $T_Y$. 
Token $T_X$ is a popular token deemed valuable and that can be exchanged into fiat currency effortlessly (e.g., USDC or WETH, which can be exchanged for USD on Coinbase~\cite{me:coinbase}).
Token $T_Y$ supports non-standard token supply control functions, like \texttt{rebase}. 

There are at least two AMM pools, denoted by $p$ and $q$. Both pools are between $T_X$ and $T_Y$. Pool $p$ has a balance-insensitive price in the sense that when $p$'s balance in token contract $T_X$ (or $T_Y$) changes, the token price in Pool $p$ does not change.
Pool $q$ has a balance-sensitive price in that Pool $q$'s token price changes along with any change of Contract Account $q$'s balance inside the token contract $T_X$ (or $T_Y$).

The adversary is a profit-driven trader, referred to as account $alice$ in this work, who monitors unconfirmed transactions in an Ethereum network and sends her own transactions to extract value. In Ethereum 2.0, $alice$ may operate as an MEV searcher within the proposer-builder separation (PBS) framework. She monitors the mempool to identify opportunities and aims to have her transactions ordered strategically relative to normal transactions, achieved through transaction pricing or by leveraging bundling services.  
In this work, $alice$ specifically targets unconfirmed victim transactions related to token supply control.

Specifically, we consider that trader $alice$ initially holds $X^a_0$ units of Token $T_X$, observes one or several transactions $\{tx\}$ related to supply control of Token $T_Y$, monitors blockchain states (e.g., the ones relevant to $p$ and $q$), and sends a transaction bundle including newly sent $\{tx'\}$ and $\{tx\}$ so that trader $alice$ owns more tokens (or the same number of tokens with higher value) after blockchain executes the bundle.

\section{Motivating Observations}
\label{sec:moex}

This section presents our motivating examples, which demonstrate how token supply control can impact extractable value.

\subsection{Code Example}
\begin{lstlisting}[language=c++,
               caption=Simplified code of AMPL token (in TokenY) and Uniswap V3 (in AMM),
               firstnumber=1,
               deletekeywords={[2]INT},
               morekeywords={clustered},
               framesep=8pt,
               xleftmargin=10pt,
               framexleftmargin=4pt,
               frame=tb,
               framerule=0pt, 
               numberstyle=\small\color{black}, 
               label=lst:t2,
               escapechar=|]
contract AMMPool {
  int reserveX, reserveY; 
  TokenY Ty; TokenX Tx;
  function swapXY(int dX){
    int dY = pricing(dX, reserveX, reserveY); |\label{lst:12:l4}|
    Ty.transfer(msg.sender, dY);
    Tx.transferFrom(msg.sender, this, dX);
    reserveX += dX; reserveY -= dY;
  }
  function pricing(int _dX, int _rX, int _rY){ |\label{lst:12:l21}|
    return (_rY*_dX)/(_rX+_dX); }} |\label{lst:12:22}|

contract TokenY {
  uint256 scale = 1e18;
  mapping(address => uint256) _baseBal; |\label{lst:12:l26}|
  function rebase(int t){scale *= t;} |\label{lst:12:l27}|
  function balanceOf(address a) returns (uint256) {
    return _baseBal[a]*scale/1e18;}|\label{lst:12:l33}|
  function transfer(address to, uint256 amount) {
    _baseBal[msg.sender] -= amount*1e18/scale;|\label{lst:12:l37}|
    _baseBal[to] += amount*1e18/scale;}}|\label{lst:12:l38}|
\end{lstlisting}

Our motivating observation stems from the real-world case where a Uniswap V3 pool~\cite{me:uniswap:v3} serves swaps between AMPL~\cite{me:ampl:token} (Token $T_Y$) and another token ($T_X$).

We present simplified code for the token and pool contracts in Listing~\ref{lst:t2}. The token contract \texttt{TokenY} includes a \texttt{rebase} function (Line~\ref{lst:12:l27}) that updates a multiplicative factor \texttt{scale}, which is then used to compute account balances in both \texttt{transfer} (Line~\ref{lst:12:l37}) and \texttt{balanceOf} (Line~\ref{lst:12:l33}).

The AMM pool contract provides a \texttt{swapXY} function that lets a caller $a$ exchange $dX$ tokens of $T_X$ for $dY$ tokens of $T_Y$. The pool computes the token price (in Line~\ref{lst:12:l4}) using a constant-product invariant (in Line~\ref{lst:12:22}) and executes the swap by calling \texttt{transfer} on both token contracts.

Notably, when determining the price in Line~\ref{lst:12:l4}, the pool contract uses the reserves of Token $T_X$ and Token $T_Y$, stored in \texttt{reserveX} and \texttt{reserveY}, respectively. The variable \texttt{reserveY} tracks the amount of $T_Y$ the pool has acquired through prior swaps. 
Importantly, {\it token reserves differ from token balances}: 

$$
\texttt{reserveY} \neq \texttt{Ty.balanceOf(p)}
$$

where $p$ denotes the address of the AMM pool contract.
If tokens are sent directly to the pool account $p$ without involving swaps in the pool contract, \texttt{reserveY} remains unchanged and the token price is unaffected. For instance, a call to \texttt{Ty.rebase($t$)} updates \texttt{Ty.balanceOf(p)} but not \texttt{reserveY}. 
We refer to this property as {\it price insensitivity}.

\subsection{\cMEV from Code Example}
\label{sec:moex:cmev}

We begin with a common value-extraction template: a sandwich attack, where trader $alice$ executes a front-running transaction $tx_0$ and a back-running transaction $tx_2$, wrapping a victim transaction $tx_1$. Specifically, trader $alice$, initially holding token $T_X$, first sends $tx_0$ to swap $T_X$ for $T_Y$ on Pool $p$. The victim's transaction $tx_1$ then executes an operation, which varies based on different cases described next. Finally, $alice$ sends $tx_2$ to swap $T_Y$ back into $T_X$ on the same pool $p$.

\noindent{\bf Case B0 (Sandwiching swap)}: 
In the standard sandwich attack, the victim transaction $tx_1$ is typically a whale or large token swap on the same pool $p$, causing a significant price impact. We denote this strategy as B0, showing the sequence of three transactions over time.
Strategy B0 is profitable because $tx_1$ shifts the token price, allowing $alice$ to buy $T_Y$ at a lower price in $tx_0$ and sell it at a higher price in $tx_2$. The root cause is that the victim's \texttt{swap} operation in $tx_1$ directly influences the token price within the DEX pool $p$.

\noindent{\bf Case B1 (Sandwiching rebase)}:
Directly applying the sandwich attack template to a victim transaction $tx_1$ that performs a \texttt{rebase} operation does not yield profit.

In Case B1, we retain $tx_0$ and $tx_2$ from B0 but replace the whale swap in $tx_1$ with a \texttt{rebase} operation on token $T_Y$. The sequence of transactions is illustrated in Figure~\ref{fig:mev:template:0}.

To assess profitability, we consider a special case where the pool $p$ follows a constant-product market maker (CPMM) model, deriving its price from token balances, and where \texttt{rebase} operates multiplicatively.
Our back-of-the-envelope calculation confirms that B1 is unprofitable. The reason is that while the victim \texttt{rebase} operation influences the token price and initially appears to enable a buy-low-sell-high strategy like B0, the apparent profit is negated because trader $alice$'s $T_Y$ holdings are also rebased after $tx_2$. In other words, \texttt{rebase} affects both $alice$'s tokens and the pool's exchange price, causing the two effects to cancel out, resulting in zero profit.

\noindent{\bf Case D1+ (Sandwiching rebase w. price-insensitive pools)}: We consider a scenario where the pool's price remains unaffected by the rebase operation. That is, while rebase changes both trader $alice$'s and pool $p$'s holdings of token $T_Y$, it does not influence the exchange price. Such price-insensitive cases exist in real-world smart contracts, including Uniswap V3.

Now, we re-run the sandwich strategy from Case B1, but with a price-insensitive pool, denoted by $q$. This new case, D1+ in Figure~\ref{fig:mev:template:0}, re-enables profitability by avoiding the price-token dependency that previously canceled out the profit. Specifically, when rebase increases the token supply, trader $alice$'s holdings of $T_Y$ increase after $tx_1$. Since pool $q$ maintains a fixed exchange price, $alice$ can swap the additional $T_Y$ for $T_X$ without price impact, ultimately acquiring more tokens.

\begin{center}\fbox{\parbox{0.90\linewidth}{
\bf{Observation}: {\it 
Profitability can arise from two previously unexamined sources: (1) token supply control that alters the balances of both pool and trader accounts, and (2) exchange prices that remain insensitive to some forms of token balance updates, including those induced by supply control.
} 
}}\end{center}

\subsection{(In)feasibility of Existing Approaches}
\label{sec:relatedwork}
\label{sec:relatedwork:1}


\noindent{\bf 
Nyx~\cite{DBLP:conf/sp/ZhangZSLWLZC24}} defines a class of profitable frontrunning vulnerabilities by refining state-inconsistent bugs~\cite{DBLP:conf/sp/Bose0C0KV22} to account for financial impact. At a high level, its bug oracle captures cases where, given the same initial contract state, reordering function calls leads to inconsistent final states, switching among which an attacker can extract value from a victim account.

To discover such vulnerabilities, Nyx systematically explores function pairs that satisfy its oracle. It starts by statically analyzing all function pairs in the target smart contracts, selecting those in which invoking one function influences the token transfer behavior inside the other (specifically, where a token transfer is reachable in the second function and the transfer amount is controllable by the first function). These candidate pairs are then subjected to more computationally intensive symbolic validation to confirm the presence of exploitable flaws.


{\color{blue}
While the flaws targeted in this work, such as D1+, conform to the state inconsistency bug oracle, Nyx's static analysis cannot detect them. This limitation stems from Nyx's design choice to focus only on application-layer contracts.

Specifically, Nyx detects function pairs where invoking the first directly influences the amount of tokens transferred by the second. It performs static analysis on application-layer contracts, such as pools, to identify read-after-write data dependencies across functions.
This approach suffices for cases like B0, where transaction $tx_2$ updates the pool contract $p$'s state, such as token reserves used for price calculation, which is later read by $tx_3$.
However, Nyx cannot detect cases like D1+, where the dependency between $tx_2$ and $tx_3$ manifests at runtime through a call argument, such as $aY$ in $tx_3$, and spans two layers, with $tx_2$ invoking a token-layer contract and $tx_3$ invoking an application-layer contract. Because Nyx relies on static analysis within a single layer, it cannot capture this cross-layer, runtime dependency.
}

\definecolor{mygreen}{rgb}{0,0.6,0}

\begin{lstlisting}[language=c++,
               caption={Pool smart contracts with price-dependent and independent \texttt{rebaseY} functions for evaluation by Nyx},
               firstnumber=1,
               deletekeywords={[2]INT},
               morekeywords={clustered},
               framesep=8pt,
               xleftmargin=10pt,
               framexleftmargin=4pt,
               frame=tb,
               framerule=0pt, 
               numberstyle=\small\color{black}, 
               label=lst:amm:nyx,
               escapechar=|,
               escapeinside={(*}{*)}]
contract AMMPool {
  ERC20 tokenX,tokenY; address vault;
  uint256 public reserveX,reserveY;
  function swapYX(uint dY) public payable {
    uint256 c = reserveX * reserveY;
    uint256 dX = reserveX - c / (reserveY + dY);
    tokenY.transferFrom(msg.sender, this, dY);
    tokenX.transfer(msg.sender, dX);
    reserveY -= dY; reserveX += dX;
  }
  function rebaseY-p(uint256 factor) public payable {
    uint256 bal_sender;
    bal_sender = tokenY.balanceOf(msg.sender);
    tokenY.transferFrom(vault, msg.sender, factor*bal_sender);
    tokenY.transferFrom(vault, this, factor*reserveY);
    (*\textbf{reserveY+=factor*reserveY;}*) //only in price-sensitive pool(*\label{line:diffline}*)
  }
  function rebaseY-q(uint256 factor) public payable {
    uint256 bal_sender;
    bal_sender = tokenY.balanceOf(msg.sender);
    tokenY.transferFrom(vault, msg.sender, factor*bal_sender);
    tokenY.transferFrom(vault, this, factor*reserveY);
}}

\end{lstlisting}

\noindent{\bf 
Evaluation on Nyx}: 
To observe Nyx's ability in detecting Example D1+ (in \S~\ref{sec:moex:cmev}), a naive approach is to directly feed the related smart contracts in Listing~\ref{lst:t2} into Nyx's open-source tool~\cite{me:nyx:tool}. However, this does not work, engineering-wise: The Nyx implementation supports only application-layer contracts, and the token-layer contracts required in Example D1+ (i.e., where the \texttt{rebase} function resides) do not match Nyx's expected interface.

To adapt our example for Nyx, we refactor the smart-contract code as shown in Listing~\ref{lst:amm:nyx}. Specifically, the \texttt{rebase} function is ``uplifted'' into the pool layer and internally uses \texttt{transfer} to distribute rebased tokens to two accounts (the pool contract address $p$ and the attacker account $a$).

We implement both a price-dependent variant, \texttt{rebaseY-p}, and a price-independent variant, \texttt{rebaseY-q}. For fair comparison, the two variants differ by {\it only one line of code}: the price-dependent variant updates token reserves in Line~\ref{line:diffline} of Listing~\ref{lst:amm:nyx}, while the price-independent one omits the line.

We construct Case D1+ by transaction sequence: 
$tx_0$: $dY1=[a, q]$.\texttt{swapXY$(dX)$},
$tx_1$: $[b, q]$.\texttt{rebaseY-q$()$},
$tx_2$: $[a, q]$.\texttt{swapYX$(dY2)$}.
We can then analyze its profitability and derive that Case D1+ is profitable. Similarly, we can construct Case B1 (by replacing \texttt{rebaseY-q} with \texttt{rebaseY-p}) and derive it is unprofitable.

We feed both cases into Nyx~\cite{me:nyx:tool}. Nyx successfully detects the former, non-profitable case of B1 (causing a false positive), but fails to detect the latter, profitable D1+ (causing a false negative). The demo of the tool use is documented~\cite{me:tscan:demo}.

\ignore{
\noindent{\bf 
Validation analysis}:
We further validate our empirical results on the D1+ example as follows: Nyx fails to detect this opportunity because it relies on static smart-contract analysis to identify inter-transaction dependencies. However, the dependency between $tx_2$ and $tx_1$, specifically via the input $dY2$, is created only at runtime, and thus escapes Nyx's detection.

Here, searcher $a$ first swaps Token $X$ for $Y$ using pool $q$ in $tx_0$, receives additional $Y$ via Token $Y$'s \texttt{rebase} event in $tx_1$, and then swaps $Y$ back to $X$ in $tx_2$ at the original, unchanged price. The extra $Y$ tokens ($dY2 - dY1$) translate into a net profit, as confirmed in our profitability analysis in \S~\ref{sec:moex:cmev} (see Example D1+).
}

\noindent{\bf 
Foray~\cite{DBLP:conf/ccs/WenLSC0024}}
synthesizes financial attacks.
To do so, it uses static analysis to lift standard token operations into non-standard, high-level financial operations (e.g., from a pair of transfers to a token swap). It then constructs the token-flow graph with edges being those financial operations, on which cycles are found.
Cycles are used to generate constraints, where computation inside functions are manually modeled and compiled into constraints. Constraints are solved against runtime arguments to synthesize profitable transaction sequences.

While Foray can synthesize conventional MEV, it cannot synthesize the \cMEV targeted in this work: 
First, \texttt{rebase}-like functions cannot be detected by Foray's static contract analysis, which relies on standard token operations which, however, \texttt{rebase} does not invoke.
Second, profitable \cMEV patterns like D1+ do not form cycles as conventional MEV does, and cannot be captured by Foray's cycle-based methods.
Third, \cMEV that resides in a large quantity of token contracts entails automated contract analysis to scale, which Foray fails to support, as it relies on manually crafted constraints to model in-functions computation (e.g., price calculation).


\noindent{\bf 
DeFiTainter~\cite{DBLP:conf/issta/KongCWJZ23}} detects price-manipulation vulnerabilities through static contract analysis. It employs cross-contract data-flow (taint) analysis to trace connections between the data source of manipulable elements, such as function call parameters, and the data sinks in sensitive operations like \texttt{transfer} calls. However, its reliance on manual labeling of manipulable elements and sensitive operations limits the scalability of its detection capabilities.

\noindent{\bf 
SmartCAT~\cite{DBLP:conf/uss/ZhangHHMW25}}
prevents price-manipulation attacks by detecting malicious smart contracts in ``real time'', that is, immediately after deployment and before attacks occur.
The key idea is to detect patterns such as pump-and-dump and sandwich attacks, for example, $\texttt{Flashloan}(\cdot) \rightarrow \texttt{swapXY}(\cdot) \rightarrow \texttt{*}(\cdot) \rightarrow \texttt{swapYX}(\cdot)$, that manifest in attacker contracts without analyzing victim contracts such as liquidity pools.
They construct token-flow graphs from attacker contracts and ``recover'' function-call arguments to connect token flows. For instance, at the router level, the token pair in a swap call appears as arguments to router functions.


In price-manipulation attacks, all transactions in the sequence are issued by the adversary and thus can manifest within a single smart contract, i.e., the attacker's contract, as analyzed by SmartCAT. This assumption, however, does not hold for MEV transations, where an MEV sequence interleaves attacker and victim transactions and therefore do not appear in any single contract. In other words, the approaches that analyze only the attacker's smart contract, such as SmartCAT~\cite{DBLP:conf/uss/ZhangHHMW25}, are inapplicable to discovering MEV, including D1+.

\ignore{
\noindent{\bf 
SmartCAT~\cite{DBLP:conf/uss/ZhangHHMW25}} detects deployed smart contracts that facilitate price-manipulation attacks. Using known attack cases as ground truth, the authors validate their approach and identify new attack contracts in the wild. They begin by collecting call traces from attacker contracts using the Gigahorse analysis tool. Then, leveraging APIs from DEXs and flashloan providers (e.g., ERC20, Uniswap, Aave), they abstract low-level calls into high-level token flow graphs, such as $\texttt{Flashloan}(\cdot) \rightarrow \texttt{swapXY}(\cdot) \rightarrow \texttt{buyMiner}(\cdot) \rightarrow \texttt{swapYX}(\cdot)$.
SMARTCAT defines behavioral patterns for price manipulation attacks, for example, a Pump and Dump is characterized by a \texttt{swapXY} followed by a \texttt{swapYX}. If a token flow graph matches such a pattern, SMARTCAT flags the contract as a potential price-manipulation attack.

SMARTCAT performs static analysis on candidate contracts to detect known attack behaviors, such as price manipulation. In contrast, our work analyzes victim contracts to uncover previously unknown MEV flaws.
}

\noindent{\bf 
In summary},
 1) the existing research that statically smart contracts to detect data dependencies there, such as Nyx, do not capture our MEV, because the inter-transaction dependencies in our motivating examples manifest only at runtime, (i.e., between $tx_2$ and $tx_1$ via the argument $dY2$ in our example). 
2) The existing smart-contract analysis that focuses only on the application layer, including Foray, DeFiTainter, and Nyx, is ineffective, because the root cause lies in token smart contracts.
3) The existing approaches that discover price-manipulation attacks by analyzing attacker contracts do not apply to MEV discovery, because an MEV sequence consists of both attacker and victim transactions, which do not appear within a single attacker-controlled smart contract. 

Empirically, we validate Nyx's incapability of detecting our MEV, while other tools are either not open source or present engineering challenges in setup.

\ignore{
\noindent{\bf Empirical feasibility}: We tested the feasibility of existing tools, notably Nyx and Foray, in detecting our bugs. Detailed results are available online.\footnote{\url{https://sites.google.com/view/tscan2025}}
}

\section{Definitions}

The profitability of Pattern D1+ arises from two key factors:
1) The \texttt{rebase} function simultaneously alters the token balances of both the trader and the pool (i.e., $Y^a$ and $Y^p$), and
2) the exchange price observed by the backrunning \texttt{swap} remains insensitive to the pool's balance ($Y^p$), even though it is modified by \texttt{rebase}.
We formally capture these two factors by defining the concepts of token supply control functions (in \S~\ref{sec:tsc}) and price insensitivity (in \S~\ref{sec:pricedep}).

\subsection{Token Supply Control (TSC) Functions}
\label{sec:tsc}

We present two formal definitions of token-supply-control functions. We begin by defining token state transition.

\newtheorem{defn}{\bf Definition}
\begin{defn}[Token state transition]
Given a token smart contract $T_Y$, a token state refers to the set of all token holders' balances, denoted as $Y^\ast$ or $\{Y^a \mid \forall a \in A\}$, where $A$ refers to all token holders. A mechanism $f$ that triggers a transition from an initial state $Y_0^\ast$ to an end state $Y_1^\ast$ is denoted as $f: Y_0^\ast \rightarrow Y_1^\ast$.
\end{defn}

The first definition captures the general notion of token supply control as a state transition that results in a change to the total token supply. Formally,

\begin{defn}[Type-1 TSC: Change of supply] \label{def:type1-tsc}
In a token smart contract $T_Y$, a state transition $f: Y_0^\ast \rightarrow Y_1^\ast$ is a type-1 token-supply-control (TSC-1) if the following condition holds:

$$ 
\sum_{\forall{}a\in{}A^Y_0\cup{}A^Y_1} Y_0^a \neq{} \sum_{\forall{}a\in{}A^Y_0\cup{}A^Y_1} Y_1^a 
$$

Here, $A^Y_0$ (or $A^Y_1$) denotes the set of all accounts whose token balances are non-zero in $Y_0^\ast$ (or $Y_1^\ast$).
\end{defn}

In the second definition, we take a closer look at the observed Pattern D1+. A key factor driving its profitability is that the \texttt{rebase} function updates the balances of multiple accounts {\it simultaneously} and according to the same formula, for example, $Y^p_1 = Y^p_0 \cdot t$ and $Y^a_1 = Y^a_0 \cdot t$ in AMPL's \texttt{rebase}.

We formalize the second definition to capture such {\it synchronous balance updates} across multiple accounts.

\begin{defn}[Type-2 TSC: multi-account sync'ed updates]
In a token smart contract $T_Y$, a state transition $f: Y_0^\ast \rightarrow Y_1^\ast$ is a type-2 token-supply-control (TSC-2) if, for any two accounts $a$ and $b$ with non-zero balances in $Y_0^\ast$, there exists a function $g(\cdot, \cdot)$ such that

$$
\forall{}a, b\in{}Y_0\cup{}Y_1, g(Y_0^a, Y_1^a) = g(Y_0^b, Y_1^b)
$$
\end{defn}

For example, a \texttt{mint} function that updates the balance of a single account qualifies as a type-1 TSC but not a type-2 TSC.

For another example, a multiplicative \texttt{rebase}, as shown in Listing~\ref{lst:t2}, qualifies as both type-1 and type-2 TSC, since it updates the balances of multiple accounts. Specifically, \texttt{rebase} is a type-2 TSC with $f(y, t): y= y \cdot t$ and $g(Y_0, Y_1) = \frac{Y_1}{Y_0}$. It is straightforward to verify that $g(Y_0^a, Y_1^a) = t = g(Y_0^b, Y_1^b)$.

\subsection{Price-Insensitive Exchanges (PITEX)}
\label{sec:pricedep}

\begin{defn}[Token-exchange function (TEX)] \label{def:exch}
A function $\texttt{foo}$ in a deployed smart contract $p$ is said to support token exchange, or is a TEX between tokens $T_X$ and $T_Y$, if, for any account $a$ that initially hold $T_X$, invocating $[a, p].\texttt{foo}(*)$ enables the exchange of $a$'s $T_X$ tokens for $T_Y$ tokens with the pool.
\end{defn}

This abstraction captures a range of real-world DeFi services, including decentralized exchanges (DEXs), AMM pools, lending protocols, and centralized exchanges (CEXs). For instance, lending can be modeled as an exchange between a collateral token $T_X$ and a borrowed token $T_Y$. \S~\ref{sec:s2:pitex} lists a number of common and uncommon DeFi operations that we consider as token exchanges.

To define a token price's sensitivity, we focus on the {\it spot price}, the exchange rate under an infinitesimally small trade that remains constant during the trade. Here, we intentionally exclude the factor of slippage, which affects the execution price by depending on the trade size~\cite{DBLP:conf/fc/QinZLG21,me:slippage}.

\begin{figure}[!bthp]
  \centering
    \includegraphics[width=0.35\textwidth]{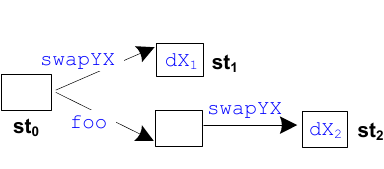}
  \caption{Defining price-insensitive exchange (PITEX).}
  \label{fig:pitex:def}
\end{figure}

Given a token-exchange pool $p$ for exchanging tokens $T_Y$ for tokens $T_X$, the pool $p$ is price-insensitive, or $p$'s spot price is insensitive to $p$'s token balance, if one can find a way to change $p$'s balance in $T_Y$ (i.e., $Y^p$) that does not affect $p$'s spot price. The formal definition is presented below, and its intuition is illustrated in Figure~\ref{fig:pitex:def}.

\begin{defn}[Price-insensitive token exchange (PITEX)]
\label{def:insensitifity}
Suppose a pool $p$ that supports a TEX function \texttt{tex} is deployed on the blockchain with initial state $st_0$.
Two function-call sequences are executed on the same $st_0$, denoted $seq_1$ and $seq_2$.
In $seq_1$, the function call is $dX_1=\texttt{tex}(dY)$.
In $seq_2$, two functions are invoked in sequence, namely $\texttt{transfer}(p)$ followed by $dX_2=\texttt{tex}(dY)$, both starting from $st_0$.

If the amount of tokens received in the first sequence equals that in the second, that is, $dX_1\equiv{}dX_2$, the TEX function is defined to be price-insensitive, or PITEX.
Details are illustrated in Equations~\ref{eqn:sensitive:difftest}.

\begin{eqnarray}
seq_1: &&
dX_1 = [b, p].\texttt{swapYX}(dY) \\
\nonumber 
seq_2: && 
[a, T_Y].\texttt{foo}(p, dY2), 
\\
\label{eqn:sensitive:difftest}
&& dX_2 = [b, p].\texttt{swapYX}(dY) \\
\nonumber
dX_1 &&\equiv dX_2
\end{eqnarray}
\end{defn}

\subsection{Defining \cMEV Templates}
\label{sec:templates}

\begin{figure}[!bthp]
  \centering
    \includegraphics[width=0.46\textwidth]{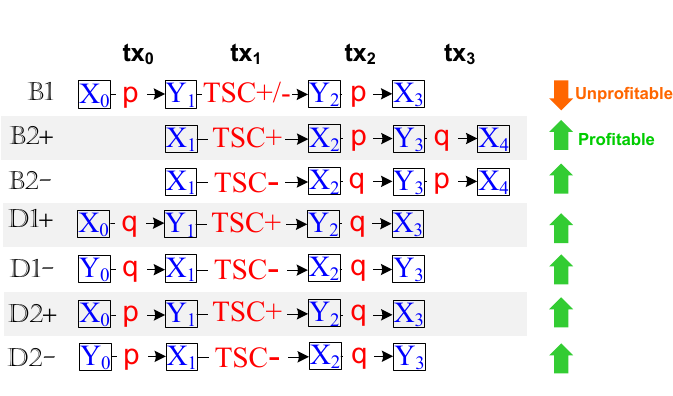}
  \caption{\cMEV templates and baselines}
  \label{fig:mev:template:0}
\end{figure}

We construct \textit{\cMEV templates} to guide the discovery and instantiation of \cMEV transactions. Unless otherwise specified, we assume transaction $tx_1$ invokes a token supply control (TSC) event on token $T_Y$.


\noindent{\it 
Templates D1 and D2}: These templates are sandwich attacks, and they follow the transaction sequence $tx_0$, $tx_1$, $tx_2$, wrapping the TSC event in $tx_1$.

\begin{itemize}[leftmargin=*]
  \item {\bf D1+/D2+ (Sandwiching positive TSC).} In $tx_0$, the trader exchanges $T_X$ for $T_Y$ using pool $p$ or $q$. In $tx_2$, the trader exchanges $T_Y$ back to $T_X$ via $q$.
  \item {\bf D1-/D2- (Sandwiching negative TSC).} In $tx_0$, the trader exchanges $T_Y$ for $T_X$ using pool $p$ or $q$. In $tx_2$, the trader exchanges $T_X$ for $T_Y$ via $q$.

D1-/D2- can be extended to a more general case with trader $a$ initially holding a third token $T_Z$. In this extended version, $tx_{-1}$ occurs before $tx_0$, and it exchanges $T_Z$ for $T_Y$ via $p$. $tx_3$ takes place after $tx_2$, and it exchanges $T_Y$ back to $T_Z$.
\end{itemize}

\noindent{\it 
Templates B2+ and B2-}: These two templates are arbitrage backrunning TSC, and they follow the transaction sequence $tx_1$, $tx_2$, $tx_3$, with the TSC event in $tx_1$ occurring first.

\begin{itemize}[leftmargin=*]
  \item {\bf B2+ (Arbitrage backrunning positive TSC).} $tx_1$ triggers a positive TSC. Then, $tx_2$ exchanges $T_X$ for $T_Y$ via pool $p$, and $tx_3$ exchanges $T_Y$ back to $T_X$ via $q$.
  \item {\bf B2- (Arbitrage backrunning negative TSC).} $tx_1$ triggers a negative TSC. Then, $tx_2$ exchanges $T_X$ for $T_Y$ via pool $q$, and $tx_3$ exchanges $T_Y$ back to $T_X$ via $p$.
\end{itemize}

All templates are illustrated in Figure~\ref{fig:mev:template:0}.

\section{The MEV-Search System}

\begin{figure}
\centering
\includegraphics[width=0.475\textwidth]{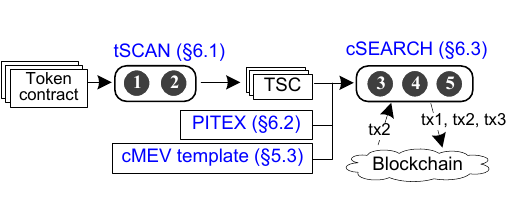}
\caption{System workflow in offline and online phases.}
\label{fig:workflow}
\end{figure}

The goal of this work is to comprehensively understand \cMEV opportunities in real-world smart contracts and to search for profitable instances against live blockchain states, ultimately materializing value that remains unexploited in current practice.

\label{sec:overview}
\noindent{\bf
System overview}: 
To this end, we propose a \cMEV-search system, as illustrated in Figure~\ref{fig:workflow}. 
Given token smart contracts deployed on a target Ethereum network (e.g., mainnet), the system first determines whether the tokens implement TSC by applying \underline{t}oken-oriented \underline{S}tatic \underline{C}ontract \underline{AN}alysis, or \staticname{} (described in \S~\ref{sec:s1:tsc}). Next, a series of target PITEX are identified from popular DeFi applications, as presented in \S~\ref{sec:s2:pitex}. The resulting PITEX pools and TSC tokens are then provided to a constraint-centric MEV search engine (\S~\ref{sec:s3:csearchr}), which continuously generates and refines constraints using both static and runtime information. Finally, these constraints are solved to synthesize concrete MEV transaction sequences.

\subsection{Detecting TSC with \staticname}
\label{sec:congene}
\label{sec:s1:tsc}
Given a token contract, \staticname statically analyzes it and determines if it's a TSC token. 
Specifically, \staticname discovers the reachable execution paths (named tPath) that can update multi-account balances, that is, TSC tokens in Definition~\ref{def:type1-tsc}.
The outcome of tPaths will be fed into the downstream operations for generating static constraints and for constraint-centric MEV searching (in \S~\ref{sec:s3:csearchr}).
Internally, the static token analysis runs in two steps: 1) generating token System Dependency Graph (tSDG) that includes intra-procedural control- and data- dependencies and inter-procedural data dependencies in the input token contracts (\ballnumber{1}), and 2) the exploration, discovery and validation of extended execution paths (tPath) that can control token supply (\ballnumber{2}).

Next, we describe the two steps in detail. We use an example token contract in List~\ref{lst:tokenanalysis:ex}.


\noindent{\bf 
tSDG construction in \ballnumber{1}}: 
Given a token contract, Algorithm~\ref{alg:tSDG} constructs a tSDG of the token that includes intra-procedural control- and data- dependencies and inter-procedural data dependencies; the produced tSDG for enabling downstream static analysis. The algorithm runs in a nested loop, where the inner loop runs intra-procedural analysis, and the outer loop pieces together the inner-loop results through inter-procedural data-flow analysis. Specifically, the core data structure in Algorithm~\ref{alg:tSDG} is a queue storing statements $q$. Initially, the queue $q$ stores all \texttt{return} statements in the \texttt{balanceOf} function of the token contract (Ln.3-7). During the analysis, the queue may store definitions of relevant state variables (as will be described). The algorithm dequeues a statement $s$ (Ln.9) in the outer loop.
Given statement $s$ in function $f$, the algorithm triggers the inner loop if $f$ has not been visited before (Ln.10-12). The inner loop constructs the intra-procedural control- and data-dependency graph of $f$ (Ln.13), upon which it further conducts backward data-flow analysis to find the data origin of $s$ in $f$. (Ln.15)

\definecolor{mygreen}{rgb}{0,0.6,0}
\lstset{ %
  backgroundcolor=\color{white},   
  basicstyle=\scriptsize\ttfamily,        
  breakatwhitespace=false,         
  breaklines=true,                 
  captionpos=b,                    
  commentstyle=\color{mygreen},    
  deletekeywords={...},            
  escapeinside={(*@}{@*)},          
  extendedchars=true,              
  keepspaces=true,                 
  keywordstyle=\color{blue},       
  language=Java,                 
  numbers=left,
  stepnumber=1,
  numbersep=5pt,                   
  numberstyle=\small\color{black}, 
  rulecolor=\color{black},         
  showspaces=false,                
  showstringspaces=false,          
  showtabs=false,                  
  stepnumber=1,                    
  stringstyle=\color{mymauve},     
  tabsize=2,                       
  caption={Example token contract},                  
  label={lst:tokenanalysis:ex},
}
\begin{figure*}[!bhtp]
\centering
\begin{minipage}{.35\textwidth}
\begin{lstlisting}
contract TokenTY {
  mapping (address => int256) balance[];
  int256 ts = 1000; int256 pause = 1; 
  int256 t1 = 0;
  TokenTY(){}
  balanceOf(address A) returns(int256){
    int256 u = balance[A];
    if(!pause)
      return u * ts;
    else return 0;}
  rebase1(){
    ts += t1;}
  rebase2(int256 t2){
    t1 = t2;}
  pauseTransfer(){
    pause = !pause; }}
\end{lstlisting}
\end{minipage}
\begin{minipage}{.625\textwidth}
\includegraphics[width=\textwidth]{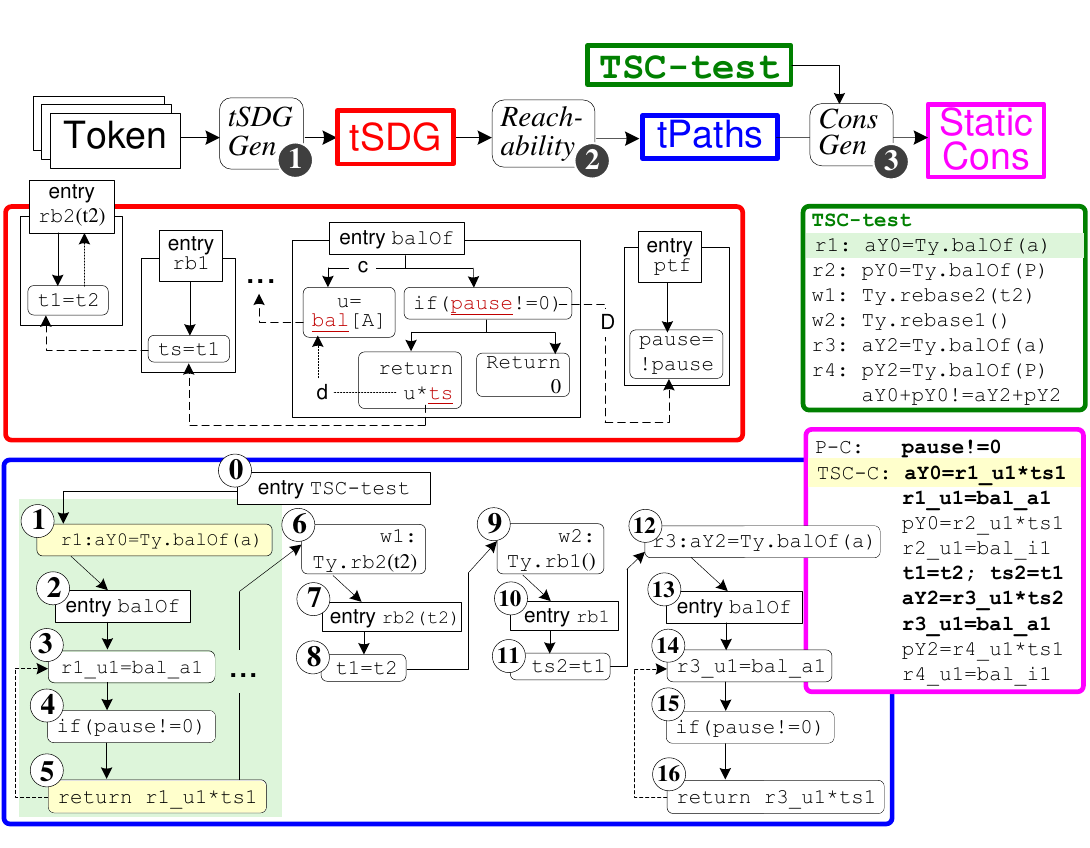}
\caption{Static token-contract analysis.}
\label{fig:interpdg:3}
\end{minipage}
\end{figure*}


\ignore{
contract TokenY {
  mapping (address => uint256) public balance[];
  int256 tSupply = 1000; int256 pause = 1;
  int256 t2 = 0;
  TokenY(){}
  balanceOf(address B) returns(uint256){
    int256 u = balance[B]; 
    if(!pause) 
      return u * tSupply;
    else return 0;}
    
  (*@\textbf{burn}@*)(address A){balance[A] -= 200/tSupply;}
  (*@\textbf{rebase1}@*)(int256 t1){tSupply *= t1;} 
  (*@\textbf{rebase2a}@*)(){tSupply += t2;}
  (*@\textbf{rebase2b}@*)(int256 t){ t2 = t;}
  (*@\textbf{rebase3}@*)(){ pause = !pause; }
}

contract SToken {
  mapping (address => uint256) public balance[];
  int256 tSupply = 1000;
  int256 pause;
  int256 v2 = 10, v3 = 20;
  SToken(){}
  balanceOf(address B) returns(uint256){
    int256 u = balance[B]; 
    if(pause != 0) 
      return u * tSupply;
    else return 0;}

  (*@\textbf{toggle}@*)(){pause = !pause;}
  (*@\textbf{burn}@*)(address A){
    balance[A] =200;}
  (*@\textbf{rebase}@*)(int256 v1){tSupply = v1 + v2 + v3;}
  (*@\textbf{rebase2}@*)(){v2 = 0;}
  (*@\textbf{rebase3}@*)(){v3 -= v2+5;}
}
}

For each data origin, if it is a state variable $v$, our system finds all the sites defining $v$ in all functions in the contract and enqueues the definitions of $v$ (Ln.16-20).
It takes no further action if the data origin is a constant or a function argument; the latter is externally controlled. Note that generating a $tSDG$ stops at the nodes that are statements referencing only state variables already considered (to prevent repeated search), constant, or function arguments.

\noindent{\bf 
TSC path discovery in \ballnumber{2}}: 
Given a token $T_Y$'s tSDG generated above, Algorithm~\ref{alg:tokenanalysis2} discovers reachable execution paths (tPaths) that control or change the token supply of $T_Y$. Internally, the algorithm runs as follows:

\noindent{\it 
2a) Data-source selection}: Given a $tSDG(T_Y)$, \staticname iterates through all \texttt{return} statements in function \texttt{balanceOf}; \staticname treats each \texttt{return} statement as a root and generates a spanning tree by following the edges in $tSDG(T_Y)$ backwardly. It then explores and finds all the tree paths that can connect the root to a tree node, which is either a function argument or a compound assignment (e.g., $v+=10$). In other words, it checks on each found tree path if there is a data flow from the data-source tree node to the root. Such a tree path can span multiple function scopes and correspond to a function-call sequence. 

For instance, in Figure~\ref{fig:interpdg:3}, on the selected candidate path highlighted in yellow, the destination is the \texttt{return} statement (Ln.9 in List~\ref{lst:tokenanalysis:ex}). The data source is the entry to function \texttt{rebase2} (or \texttt{rb2}). This data-source-to-destination path spans over three functions: \texttt{balanceOf}, \texttt{rebase1()}, and \texttt{rebase2($t2$)}. Note that there could be multiple data sources on a given destination. When selecting one tree path, we only need to include {\it one} data source (out of many), sufficient to cause the change of token supply. For instance, in Figure~\ref{fig:interpdg:3}, the same destination (Ln.9 in List~\ref{lst:tokenanalysis:ex}) is also data- or control-dependent on other variables like \texttt{u} and \texttt{pause}, and our selected tree path only tracks the data source through \texttt{ts} without \texttt{u} or \texttt{pause}.

\noindent{\it 2b) Execution path generation}: It extends each selected tree path to be an execution path. For each tree node $s$ in a function $f$, \staticname backtracks the intra-procedural control flow to reach the function entry of $f$. Then, it performs context-sensitive analysis to stitch the execution paths across multiple function scopes. This involves renaming state variables. 
In Figure~\ref{fig:interpdg:3}, the graph in the blue box depicts an execution path of three function calls, \texttt{rebase2($t2$)}, \texttt{rebase1()}, and \texttt{balanceOf}. 

Note that \staticname's inter-procedural analysis can be configured to cover multiple functions (three or more); in practice, we set this to three. In contrast, Nyx's inter-procedural analysis is limited to two functions, as its goal is to identify function pairs.
Additionally, \staticname can unroll loops up to a configurable bound $N$. To balance effectiveness and efficiency, we set $N = 1$ in practice.

\begin{algorithm}[H]
\caption{Token Analysis}
\small 
\label{alg:tSDG}
\begin{algorithmic}[1]
\Procedure{constructtSDG}{Contract $T_Y$}
\State $g0=$intraPPDG$(T_Y.\texttt{balanceOf})$;
\State $rs=g0.$findStmt$("\texttt{return}")$; 
\State $pdfDb.$add$(g0)$;
\For{$r \in rs$}
    \State $q.$enqueue$(r, \texttt{balanceOf})$;
\EndFor
\For{1}
    \State $e=q.dequeue()$;
    \If{$e == NULL \land{}$} \State {\bf break}; \EndIf
    \State $g=$intraPPDG$(e.f)$;
    \State $pdfDb.$add$(g)$;
    \State $dOrigins = $BDA$(g, e.s)$;
    \For{$d \in dOrigins$}
        \If{isStateVar$(d)$} 
          \State $dd=$findDefs$(d)$;
          \For{$df \in dd$}
             \State $q.$enqueue$(df, $findFunc$(df))$;
          \EndFor
        \EndIf
   \EndFor
\EndFor
\State \Return $pdfDb$
\EndProcedure
\end{algorithmic}
\end{algorithm}

\begin{algorithm}[H]
    \caption{Token Analysis2}\small \label{alg:tokenanalysis2}
    \begin{algorithmic}[1]
    \Procedure{constructtSDG}{Contract $T_Y$, $tSDG\_T_Y$}
    \State $g0=$intraPPDG$(T_Y.\texttt{balanceOf})$;
    \State $root=g0.$findStmt$("\texttt{return}")$;
    \State $\texttt{[}(root,node)\texttt{]}=findleaf(root,tSDG\_T_Y)$
    \For{$(r,n) \in \texttt{[}(root,node)\texttt{]}$}
        \If {isArgOrComp$(node)$}
            \State $Paths.append(path(root,node))$;
        \EndIf
    \EndFor
    \State \Return $Paths$
    \EndProcedure
      
    \end{algorithmic}
    \end{algorithm}

\begin{table*}[!hbtp] 
\caption{Characterize token exchanges across the applications of lending pools and various DEXs.}
\label{tab:exchanges:pi}
\centering\footnotesize
\begin{tabularx}{0.95\textwidth}{X|c c c|c c c c|c c c c}
& \multicolumn{3}{c|}{\textbf{\qone} in lending pools} & \multicolumn{4}{c|}{\textbf{\qtwo} in DEXs} & \multicolumn{4}{c}{\textbf{\qthree} in DEXs} \\
\cline{2-12}
         & Aave & Compound & dYdX & V2 & V3 & Balancer & Curve & V2 & V3 & Balancer & Curve \\
\hline
\hline
Token exchange & \cmark & \cmark & \cmark & \cmark & \cmark & \cmark & \cmark & \xmark & \cmark & \xmark & \xmark \\
Price insensitive & \cmark & \cmark & \cmark & \xmark & \cmark & \xmark & \xmark & NA & \ding{51}\rotatebox[origin=c]{-0}{\kern-0.6em\ding{55}} & NA & NA \\
\end{tabularx}
\end{table*}

\subsection{Identifying PITEX}
\label{sec:s2:pitex}

Table~\ref{tab:exchanges:pi} lists the token-exchanging operations that we identify with their price sensitvity. 
Specifically, we find price-insensitive exchanges in three cases: (i) borrowing and repaying tokens in lending services (\qone), (ii) token swaps in certain DEX pools such as Uniswap V3 and V4 (\qtwo), and (iii) liquidity addition and removal in Uniswap V3 and V4 (\qthree). Among these, \qone is standard, as lending operations typically occur at fixed prices, whereas \qtwo and \qthree are non-standard, since most DEXs exhibit sensitive prices.

\noindent{\bf
Borrow/repay in lending pools (\qone)}:
A well-known type of price-insensitive exchange is lending services like Aave and dYdX. As shown in Table~\ref{tab:exchanges:pi}, they maintain fixed prices, rendering them price-insensitive with respect to token balances during borrowing and repayment.

\noindent{\bf
Swaps in reserve-based DEXs (\qtwo)}:
Among AMM pools, Uniswap V3 exhibits price insensitivity. Its spot price is determined by token reserves, internal state variables updated only when the pool invokes \texttt{transfer} calls to the token contract (see Listing~\ref{lst:t2}). If someone directly calls $T_Y$'s \texttt{transfer} to send tokens to $p$ without invoking any function on $p$, the pool's balance changes, but its spot price remains unaffected.

In contrast, not all AMMs are price-insensitive. For example, Uniswap V2 calculates its spot price directly from the pool's token balances, so a direct token transfer to the pool affects the price.

\noindent{\bf
Adding liquidity to inactive ticks in DEXs (\qthree)}: 
A concentrated liquidity pool, such as Uniswap V3, partitions liquidity into discrete ranges, known as {\it ticks}. At any given time, exactly one tick is active, while the others remain inactive. Adding liquidity to the active tick follows the same mechanism as in non-concentrated pools. In contrast, adding liquidity to an inactive tick differs in that only one token, either $T_X$ or $T_Y$ but not both, is required. This process is therefore equivalent to a token swap between $T_X$ and $L$ (or $T_Y$ and $L$), where $L$ denotes the liquidity token.

Furthermore, adding liquidity to inactive ticks in concentrated pools corresponds to a fixed-price token swap, with the price determined by the slope at the tick boundary of the pool.

\subsection{Constraint-centric MEV Search (\dynamicname)}
\label{sec:s3:csearchr}

\dynamicname takes the following information as input: the \staticname outcome (i.e., tPath for TSC tokens) and our target PITEX pools. 
It uses the input to materialize the provided MEV templates and synthesizes a transaction sequence, as the output, ready to extract \cMEV value on a current or given blockchain state.

Internally, it first generates static constraints that contain symbols (Step \ballnumber{3}), then uses runtime information to refine them into dynamic constraints (Step \ballnumber{4}) before solving them for synthesis of MEV transactions (Step \ballnumber{5}).

\noindent{\bf
Static constraint generation in \ballnumber{3}}:
This stage generates feasible static constraints as follows. It extends each identified tPath in the token contract, and selects applicable MEV templates, as described in \S~\ref{sec:templates}, by exploring token-pool combinations. Given the selected templates, it finds the associated pool contracts, as described in \S~\ref{sec:s2:pitex}, and compiles them with the TSC tokens into static constraints, in which function arguments remain symbolic. The resulting constraints are stored in a watch list for dynamic online use. 
A static-constraint solver is applied to eliminate unsatisfiable instances, ensuring that only solvable static constraints are retained.

\label{sec:dynamic}

\noindent{\bf
Dynamic constraint solving}:
\dynamicname monitors unconfirmed transactions, for example through public mempools or via PBS by collaborating with private nodes such as block builders. It matches observed transactions against the TSC function sequences stored in the watch list and, for each observed TSC transaction, derives dynamic constraints (\ballnumber{4}). Specifically, the corresponding static constraints are retrieved from the watch list and instantiated using the observed function arguments and relevant blockchain states, such as token reserves in the AMM pool.

Next, the Z3 solver is used to check the solvability of the resulting dynamic constraints (\ballnumber{5}). Solving is performed iteratively, following the approach of~\cite{DBLP:conf/ccs/BabelJ0KKJ23}, to identify solutions that maximize extracted value. Each observed TSC-triggering transaction is evaluated against multiple \cMEV templates, and the template that yields the highest extractable value, determined through a tournament-style comparison, is selected to generate the exploit transaction.

\vspace{-0.1in}
\section{Conclusion}
This paper tackles the detection and searching of \cMEV, a new class of MEV opportunities arising from token supply control and price-insensitive exchanges. Evaluation shows that \dynamicname{} can extract up to $10\times$ more profit than current MEV searchers.

\section*{Acknowledgments}
The authors thank Jin Yang for conducting experiments on an earlier design of this work. 
This work was partially supported by two academic grants from the Ethereum Foundation and by NSF grants CNS-2139801, CNS-1815814, and DGE-2104532.

\clearpage

\section*{Ethical Considerations}

This research has implications for multiple {\bf stakeholders} in decentralized finance ecosystems, including block validators, DEX swappers, particularly liquidity providers participating in price-insensitive pools with negative TSC functions, and MEV searchers operating on DEXs. 

Greater MEV opportunities have a significant {\bf impact}, encouraging more transactions and making swaps more active. Validators benefit from increased transaction fees. Users are only unable to trade when the pool is locked, so they are not significantly impacted. However, our findings indicate that adding liquidity to such insensitive pools may expose providers to previously underappreciated security risks, while also potentially increasing extractable MEV.

Recognizing the sensitivity of these outcomes, we took measures to {\bf mitigate} real-world impact. All analyses and measurements were conducted through simulations on historical blockchain data, ensuring no interference with live protocols or users' assets. We adhered to responsible disclosure practices by reporting the identified vulnerabilities to the Uniswap community, and the relevant teams acknowledged the issues. To prevent misuse, we have not released exploit code or any artifacts that could facilitate attacks. 

Our {\bf intent} in conducting and publishing this research is to raise awareness of structural security weaknesses in price-insensitive pools and to contribute constructive insights that support the development of safer and more resilient DeFi protocols.

\bibliographystyle{plain}
\bibliography{bkc,yuzhetang,ads,lsm,odb,cacheattacks,sc,crypto,sgx,diffpriv,txtbk,distrkvs,vc}

@inproceedings{DBLP:conf/sp/Bose0C0KV22,
  author       = {Priyanka Bose and
                  Dipanjan Das and
                  Yanju Chen and
                  Yu Feng and
                  Christopher Kruegel and
                  Giovanni Vigna},
  title        = {{SAILFISH:} Vetting Smart Contract State-Inconsistency Bugs in Seconds},
  booktitle    = {43rd {IEEE} Symposium on Security and Privacy, {SP} 2022, San Francisco,
                  CA, USA, May 22-26, 2022},
  pages        = {161--178},
  publisher    = {{IEEE}},
  year         = {2022},
  url          = {https://doi.org/10.1109/SP46214.2022.9833721},
  doi          = {10.1109/SP46214.2022.9833721},
  bibsource    = {dblp computer science bibliography, https://dblp.org}
}

@inproceedings{DBLP:conf/ccs/WenLSC0024,
  author       = {Hongbo Wen and
                  Hanzhi Liu and
                  Jiaxin Song and
                  Yanju Chen and
                  Wenbo Guo and
                  Yu Feng},
  editor       = {Bo Luo and
                  Xiaojing Liao and
                  Jun Xu and
                  Engin Kirda and
                  David Lie},
  title        = {{FORAY:} Towards Effective Attack Synthesis against Deep Logical Vulnerabilities
                  in DeFi Protocols},
  booktitle    = {Proceedings of the 2024 on {ACM} {SIGSAC} Conference on Computer and
                  Communications Security, {CCS} 2024, Salt Lake City, UT, USA, October
                  14-18, 2024},
  pages        = {1001--1015},
  publisher    = {{ACM}},
  year         = {2024},
  url          = {https://doi.org/10.1145/3658644.3690293},
  doi          = {10.1145/3658644.3690293},
  bibsource    = {dblp computer science bibliography, https://dblp.org}
}

@misc{me:nyx:tool,
author = {nyx},
title = {Nyx: Detecting Exploitable Front-Running Vulnerabilities in Smart Contracts},
year = {2025},
howpublished = {\url{https://github.com/Troublor/Nyx/tree/main}}}

@misc{me:erc20:api,
   author = {ERC20-api},
   title = {OpenZeppelin's ERC20 token interfaces},
   howpublished = {\url{https://github.com/OpenZeppelin/openzeppelin-contracts/blob/master/contracts/token/ERC20/IERC20.sol}}
}

@inproceedings{DBLP:conf/ccs/LiLHLWNYCC23,
  author       = {Zihao Li and
                  Jianfeng Li and
                  Zheyuan He and
                  Xiapu Luo and
                  Ting Wang and
                  Xiaoze Ni and
                  Wenwu Yang and
                  Xi Chen and
                  Ting Chen},
  editor       = {Weizhi Meng and
                  Christian Damsgaard Jensen and
                  Cas Cremers and
                  Engin Kirda},
  title        = {Demystifying DeFi {MEV} Activities in Flashbots Bundle},
  booktitle    = {Proceedings of the 2023 {ACM} {SIGSAC} Conference on Computer and
                  Communications Security, {CCS} 2023, Copenhagen, Denmark, November
                  26-30, 2023},
  pages        = {165--179},
  publisher    = {{ACM}},
  year         = {2023},
  doi          = {10.1145/3576915.3616590},
  bibsource    = {dblp computer science bibliography, https://dblp.org}
}

@inproceedings{DBLP:conf/sp/DaianGKLZBBJ20,
  author    = {Philip Daian and
               Steven Goldfeder and
               Tyler Kell and
               Yunqi Li and
               Xueyuan Zhao and
               Iddo Bentov and
               Lorenz Breidenbach and
               Ari Juels},
  title     = {Flash Boys 2.0: Frontrunning in Decentralized Exchanges, Miner Extractable
               Value, and Consensus Instability},
  booktitle = {2020 {IEEE} Symposium on Security and Privacy, {SP} 2020, San Francisco,
               CA, USA, May 18-21, 2020},
  pages     = {910--927},
  publisher = {{IEEE}},
  year      = {2020},
  url       = {https://doi.org/10.1109/SP40000.2020.00040},
  doi       = {10.1109/SP40000.2020.00040},
  bibsource = {dblp computer science bibliography, https://dblp.org}
}

@INPROCEEDINGS{DBLP:conf/sp/ZhangZSLWLZC24,
  author={Zhang, Wuqi and Zhang, Zhuo and Shi, Qingkai and Liu, Lu and Wei, Lili and Liu, Yepang and Zhang, Xiangyu and Cheung, Shing-Chi},
  booktitle={2024 IEEE Symposium on Security and Privacy (SP)}, 
  title={Nyx: Detecting Exploitable Front-Running Vulnerabilities in Smart Contracts}, 
  year={2024},
  volume={},
  number={},
  pages={2198-2216},
  doi={10.1109/SP54263.2024.00146}}

@misc{me:slither,
  key = {Slither},
  title = {Slither},
  howpublished = {\url{https://github.com/crytic/slither}},
  year = {2024}
}

@inproceedings{DBLP:conf/sp/ZhouQCLG21,
  author    = {Liyi Zhou and
               Kaihua Qin and
               Antoine Cully and
               Benjamin Livshits and
               Arthur Gervais},
  title     = {On the Just-In-Time Discovery of Profit-Generating Transactions in
               DeFi Protocols},
  booktitle = {42nd {IEEE} Symposium on Security and Privacy, {SP} 2021, San Francisco,
               CA, USA, 24-27 May 2021},
  pages     = {919--936},
  publisher = {{IEEE}},
  year      = {2021},
  url       = {https://doi.org/10.1109/SP40001.2021.00113},
  doi       = {10.1109/SP40001.2021.00113},
  bibsource = {dblp computer science bibliography, https://dblp.org}
}

@inproceedings{DBLP:conf/www/WangCWZDW22,
  author       = {Ye Wang and
                  Yan Chen and
                  Haotian Wu and
                  Liyi Zhou and
                  Shuiguang Deng and
                  Roger Wattenhofer},
  editor       = {Fr{\'{e}}d{\'{e}}rique Laforest and
                  Rapha{\"{e}}l Troncy and
                  Elena Simperl and
                  Deepak Agarwal and
                  Aristides Gionis and
                  Ivan Herman and
                  Lionel M{\'{e}}dini},
  title        = {Cyclic Arbitrage in Decentralized Exchanges},
  booktitle    = {Companion of The Web Conference 2022, Virtual Event / Lyon, France,
                  April 25 - 29, 2022},
  pages        = {12--19},
  publisher    = {{ACM}},
  year         = {2022},
  url          = {https://doi.org/10.1145/3487553.3524201},
  doi          = {10.1145/3487553.3524201},
  bibsource    = {dblp computer science bibliography, https://dblp.org}
}

@inproceedings{DBLP:conf/uss/TorresCS21,
  author    = {Christof Ferreira Torres and
               Ramiro Camino and
               Radu State},
  editor    = {Michael Bailey and
               Rachel Greenstadt},
  title     = {Frontrunner Jones and the Raiders of the Dark Forest: An Empirical
               Study of Frontrunning on the Ethereum Blockchain},
  booktitle = {30th {USENIX} Security Symposium, {USENIX} Security 2021, August 11-13,
               2021},
  pages     = {1343--1359},
  publisher = {{USENIX} Association},
  year      = {2021},
  url       = {https://www.usenix.org/conference/usenixsecurity21/presentation/torres},
  bibsource = {dblp computer science bibliography, https://dblp.org}
}

@misc{me:coinbase,
  author = {coinbase},
  title = {coinbase},
  year = {2012},
  howpublished = {\url{https://www.coinbase.com/explore}},
  urldate = {2024-12-03}
}

@misc{me:ampl:token, 
   author = {AMPL},
   title = {Ampleforth (AMPL) token},
   year = {2025},
   howpublished = {\url{https://www.ampleforth.org/}}
}

@misc{me:slippage,
   author = {slippage},
   title = {Slippage definition \& example.},
   howpublished = {\url{https://www.investopedia.com/terms/s/slipp age.asp}}
}

@misc{me:uniswap:v3,
  author = {uniswap:v3},
  title = {Source code of Uniswap V3},
  year = {2025},
  howpublished = {\url{https://github.com/Uniswap/v3-core}},
  urldate = {2025-04-22}
}

@misc{me:uniswapLP,
  author = {uniswapLP},
  title = {Source code of Uniswap LP token on Github: UniswapV2ERC20.sol},
  year = {2025},
  howpublished = {\url{https://github.com/Uniswap/v2-core/blob/master/contracts/UniswapV2ERC20.sol#L30}},
  urldate = {2025-04-22}
}

@inproceedings{DBLP:conf/ccs/BabelJ0KKJ23,
  author       = {Kushal Babel and
                  Mojan Javaheripi and
                  Yan Ji and
                  Mahimna Kelkar and
                  Farinaz Koushanfar and
                  Ari Juels},
  editor       = {Weizhi Meng and
                  Christian Damsgaard Jensen and
                  Cas Cremers and
                  Engin Kirda},
  title        = {Lanturn: Measuring Economic Security of Smart Contracts Through Adaptive
                  Learning},
  booktitle    = {Proceedings of the 2023 {ACM} {SIGSAC} Conference on Computer and
                  Communications Security, {CCS} 2023, Copenhagen, Denmark, November
                  26-30, 2023},
  pages        = {1212--1226},
  publisher    = {{ACM}},
  year         = {2023},
  url          = {https://doi.org/10.1145/3576915.3623204},
  doi          = {10.1145/3576915.3623204},
  bibsource    = {dblp computer science bibliography, https://dblp.org}
}

@inproceedings{DBLP:conf/issta/KongCWJZ23,
  author       = {Queping Kong and
                  Jiachi Chen and
                  Yanlin Wang and
                  Zigui Jiang and
                  Zibin Zheng},
  editor       = {Ren{\'{e}} Just and
                  Gordon Fraser},
  title        = {DeFiTainter: Detecting Price Manipulation Vulnerabilities in DeFi
                  Protocols},
  booktitle    = {Proceedings of the 32nd {ACM} {SIGSOFT} International Symposium on
                  Software Testing and Analysis, {ISSTA} 2023, Seattle, WA, USA, July
                  17-21, 2023},
  pages        = {1144--1156},
  publisher    = {{ACM}},
  year         = {2023},
  url          = {https://doi.org/10.1145/3597926.3598124},
  doi          = {10.1145/3597926.3598124},
  bibsource    = {dblp computer science bibliography, https://dblp.org}
}

@inproceedings{DBLP:conf/fc/QinZLG21,
  author       = {Kaihua Qin and
                  Liyi Zhou and
                  Benjamin Livshits and
                  Arthur Gervais},
  editor       = {Nikita Borisov and
                  Claudia D{\'{\i}}az},
  title        = {Attacking the DeFi Ecosystem with Flash Loans for Fun and Profit},
  booktitle    = {Financial Cryptography and Data Security - 25th International Conference,
                  {FC} 2021, Virtual Event, March 1-5, 2021, Revised Selected Papers,
                  Part {I}},
  series       = {Lecture Notes in Computer Science},
  volume       = {12674},
  pages        = {3--32},
  publisher    = {Springer},
  year         = {2021},
  url          = {https://doi.org/10.1007/978-3-662-64322-8\_1},
  doi          = {10.1007/978-3-662-64322-8\_1},
  bibsource    = {dblp computer science bibliography, https://dblp.org}
}

@inproceedings{DBLP:conf/sp/QinZG22,
  author       = {Kaihua Qin and
                  Liyi Zhou and
                  Arthur Gervais},
  title        = {Quantifying Blockchain Extractable Value: How dark is the forest?},
  booktitle    = {43rd {IEEE} Symposium on Security and Privacy, {SP} 2022, San Francisco,
                  CA, USA, May 22-26, 2022},
  pages        = {198--214},
  publisher    = {{IEEE}},
  year         = {2022},
  url          = {https://doi.org/10.1109/SP46214.2022.9833734},
  doi          = {10.1109/SP46214.2022.9833734},
  bibsource    = {dblp computer science bibliography, https://dblp.org}
}

@misc{zhang2025followingdevilsfootprintrealtime,
      title={Following Devils' Footprint: Towards Real-time Detection of Price Manipulation Attacks}, 
      author={Bosi Zhang and Ningyu He and Xiaohui Hu and Kai Ma and Haoyu Wang},
      year={2025},
      eprint={2502.03718},
      archivePrefix={arXiv},
      primaryClass={cs.CR},
      url={https://arxiv.org/abs/2502.03718}, 
}

@inproceedings{DBLP:conf/uss/McLaughlinKV23,
  author       = {Robert McLaughlin and
                  Christopher Kruegel and
                  Giovanni Vigna},
  editor       = {Joseph A. Calandrino and
                  Carmela Troncoso},
  title        = {A Large Scale Study of the Ethereum Arbitrage Ecosystem},
  booktitle    = {32nd {USENIX} Security Symposium, {USENIX} Security 2023, Anaheim,
                  CA, USA, August 9-11, 2023},
  pages        = {3295--3312},
  publisher    = {{USENIX} Association},
  year         = {2023},
  url          = {https://www.usenix.org/conference/usenixsecurity23/presentation/mclaughlin},
  bibsource    = {dblp computer science bibliography, https://dblp.org}
}

@inproceedings{DBLP:conf/uss/0002LM0023,
  author       = {Zhuo Zhang and
                  Zhiqiang Lin and
                  Marcelo Morales and
                  Xiangyu Zhang and
                  Kaiyuan Zhang},
  editor       = {Joseph A. Calandrino and
                  Carmela Troncoso},
  title        = {Your Exploit is Mine: Instantly Synthesizing Counterattack Smart Contract},
  booktitle    = {32nd {USENIX} Security Symposium, {USENIX} Security 2023, Anaheim,
                  CA, USA, August 9-11, 2023},
  pages        = {1757--1774},
  publisher    = {{USENIX} Association},
  year         = {2023},
  url          = {https://www.usenix.org/conference/usenixsecurity23/presentation/zhang-zhuo-exploit},
  bibsource    = {dblp computer science bibliography, https://dblp.org}
}

@inproceedings{DBLP:conf/uss/QinCZLSG23,
  author       = {Kaihua Qin and
                  Stefanos Chaliasos and
                  Liyi Zhou and
                  Benjamin Livshits and
                  Dawn Song and
                  Arthur Gervais},
  editor       = {Joseph A. Calandrino and
                  Carmela Troncoso},
  title        = {The Blockchain Imitation Game},
  booktitle    = {32nd {USENIX} Security Symposium, {USENIX} Security 2023, Anaheim,
                  CA, USA, August 9-11, 2023},
  pages        = {3961--3978},
  publisher    = {{USENIX} Association},
  year         = {2023},
  url          = {https://www.usenix.org/conference/usenixsecurity23/presentation/qin},
  bibsource    = {dblp computer science bibliography, https://dblp.org}
}

@inproceedings{DBLP:conf/uss/ZhangHHMW25,
  author       = {Bosi Zhang and
                  Ningyu He and
                  Xiaohui Hu and
                  Kai Ma and
                  Haoyu Wang},
  editor       = {Lujo Bauer and
                  Giancarlo Pellegrino},
  title        = {Following Devils' Footprint: Towards Real-time Detection of Price
                  Manipulation Attacks},
  booktitle    = {34th {USENIX} Security Symposium, {USENIX} Security 2025, Seattle,
                  WA, USA, August 13-15, 2025},
  pages        = {4127--4145},
  publisher    = {{USENIX} Association},
  year         = {2025},
  url          = {https://www.usenix.org/conference/usenixsecurity25/presentation/zhang-bosi},
  bibsource    = {dblp computer science bibliography, https://dblp.org}
}

@inproceedings{DBLP:conf/sp/BabelDKJ23,
  author       = {Kushal Babel and
                  Philip Daian and
                  Mahimna Kelkar and
                  Ari Juels},
  title        = {Clockwork Finance: Automated Analysis of Economic Security in Smart
                  Contracts},
  booktitle    = {44th {IEEE} Symposium on Security and Privacy, {SP} 2023, San Francisco,
                  CA, USA, May 21-25, 2023},
  pages        = {2499--2516},
  publisher    = {{IEEE}},
  year         = {2023},
  url          = {https://doi.org/10.1109/SP46215.2023.10179346},
  doi          = {10.1109/SP46215.2023.10179346},
  bibsource    = {dblp computer science bibliography, https://dblp.org}
}

@misc{me:tscan:demo,
  author    = {Anonymized authors},
title = {Feasibility of Existing Tools in Detecting tSCAN Bugs},
howpublished = {\url{https://sites.google.com/view/tscan2025/}}
}

\clearpage

\end{document}